\documentclass[letterpaper]{article}

\usepackage[T1]{fontenc}

\usepackage{geometry}
\usepackage{setspace}

\usepackage[style = chem-acs, articletitle = true, doi = true]{biblatex}

\DeclareBibliographyDriver{inproceedings}{%
  \usebibmacro{bibindex}%
  \usebibmacro{begentry}%
  \usebibmacro{author/translator+others}%
  \setunit{\labelnamepunct}\newblock
  \iftoggle{bbx:articletitle}
    {%
      \usebibmacro{title}%
      \printunit{\adddot\addspace}%
    }
    {}%
  \usebibmacro{byauthor}%
  \setunit{\addspace}%
  \usebibmacro{in:}%
  \usebibmacro{maintitle+booktitle}%
  \newunit\newblock
  \usebibmacro{event+venue+date}%
  \newunit\newblock
  \usebibmacro{byeditor+others}%
  \newunit\newblock
  \printfield{volumes}%
  \setunit{\addsemicolon\space}%
  \printfield{note}%
  \newunit\newblock
  \usebibmacro{publisher+location+date}%
  \setunit{\addsemicolon\space}%
  \printfield{volume}%
  \setunit{\addsemicolon\space}%
  \usebibmacro{chapter+pages}%
  \newunit\newblock
  \iftoggle{bbx:isbn}
    {\printfield{isbn}}
    {}%
  \newunit\newblock
  \usebibmacro{doi+eprint+url}%
  \newunit\newblock
  \usebibmacro{addendum+pubstate}%
  \setunit{\bibpagerefpunct}\newblock
  \usebibmacro{pageref}%
  \newunit\newblock
  \iftoggle{bbx:related}
    {\usebibmacro{related:init}%
     \usebibmacro{related}}
    {}%
  \usebibmacro{finentry}%
}

\DeclareFieldFormat[inproceedings]{title}{#1}

\usepackage{graphicx}
\usepackage{float}
\newfloat{scheme}{htbp}{los}
\floatname{scheme}{Scheme}
\floatname{chart}{Chart}
\newfloat{graph}{htbp}{loh}

\usepackage{textgreek}
\usepackage{tabularray}
\usepackage{amsmath}
\usepackage{amssymb}
\usepackage{hyperref}

\usepackage{chemformula} 
\usepackage[version = 4]{mhchem} 

\usepackage{authblk}
\author[1,*,§]{Jared Benson}
\author[1,*]{Sanghyeok Park}
\author[1]{Owen M. Eskandari}
\author[1]{M. A. Wolfe}
\author[1]{Brighton X. Coe}
\author[1,**]{J. P. Dodson}
\author[2]{S. N. Coppersmith}
\author[1]{Mark Friesen}
\author[1]{M. A. Eriksson}

\affil[1]{Department of Physics, University of Wisconsin-Madison, Madison, WI 53706, USA}
\affil[2]{School of Physics, University of New South Wales, Sydney, NSW 2052, Australia}
\affil[*]{These authors contributed equally to this work.}
\affil[**]{Present address: Q-CTRL, Los Angeles, California, USA}

\title{Individually tunable Si/SiGe quantum dot operating voltages via gate-biased illumination}
\date{\textsuperscript{§}Email: jcbenson2@wisc.edu}

\begin{document}

\maketitle

\begin{abstract}
Semiconductor quantum dot qubits often require very different voltages on each gate to bring them to a correct operating point. Here, we present a method by which one can controllably and repeatably alter the nanoscale trapped charge distribution at an oxide-semiconductor interface. We demonstrate this method on a Si/SiGe quantum dot device, and we find that the operating voltages can be controlled and made much more uniform. The method relies on illumination with near-infrared light in the presence of applied gate voltages, and it enables the tuning of the device operating point on a gate-by-gate basis. We present an explanation of the underlying physics using self-consistent Schr{\"o}dinger-Poisson simulations. As an application of this method, we tune a triple quantum dot to have uniform and small operating voltages in the (1,1,1) charge configuration. Importantly, we show that shifting the operating voltages in this way does not change the measured charge noise.
\end{abstract}


\section{Introduction}

Gate-defined semiconductor quantum dots offer a compelling candidate platform for quantum computing due to their small footprint and compatibility with conventional nanofabrication techniques~\cite{Ansaloni2020-av, Zwerver2022-vj, George2025-fu, Koch2025-kp, Huckemann2025-nn, Steinacker2025-uv}. However, typical device design parameters can lead to inconsistent and large operating voltages, in part due to charge traps at the interface between the semiconductor heterostructure and the gate oxide layer~\cite{Neyens2024-lj, Loenders2026-zm}. This can lead to difficult device tuning and incompatibility with instrumentation, particularly with cryo-CMOS electronics that tend to have severe power constraints~\cite{Pauka2021-lh, Park2021-fp, Mouny2023-aq, Schreckenberg2023-zr, Bartee2025-oh, Dumoulin-Stuyck2026-yh}. These problems are relevant for scaling, where uniformity and cryogenic control are especially important~\cite{Vandersypen2017-kd, Veldhorst2017-uf, Kunne2024-pu}. While this can be mitigated through improved fabrication process control, this is a hard problem to overcome. Moreover, increasing the uniformity of the gates through better fabrication may not be a complete solution given that different types of gates, which are designed to fulfill different roles within the functioning of quantum dots, tend to be operated at substantially different voltages. An alternative approach to solve these issues is to improve the device characteristics \emph{in-situ} after it is fabricated and cooled to cryogenic temperatures.

Here, we present a technique to control the density of trapped charge underneath nanoscale gates in an oxide-semiconductor device, like those that define quantum dots, thereby enabling tunable control over the gate voltages required for device operation. The technique, which we call gate-biased illumination, is general and involves illuminating the device with above bandgap near-infrared light while simultaneously applying precisely chosen voltages to each gate. Compared to other methods that have been demonstrated to adjust the operating voltages of quantum devices, such as bias stressing~\cite{Meyer2023-ny, Meyer2023-us, Massai2024-gv} and bias cooling~\cite{Ferrero2024-kb, Diebel2025-gm}, gate-biased illumination is fast, and it allows for more control over exactly where charge is rearranged in the device. We can measure the threshold for gate voltages above which charge is injected into the oxide layer, and we take care to avoid this limit~\cite{Wolfe2024-hs}. 

We demonstrate that gate-biased illumination enables shifting the operating voltages on a gate-by-gate basis in a repeatable manner. As an application, we show how this technique enables placing the quantum dots in a low-gate-voltage operation regime and increases the uniformity of the required gate voltages. To explain the mechanism that makes it possible to tune the operating voltages, we simulate the process and the resulting pinch-off voltage shifts using Schr{\"o}dinger-Poisson simulations. These shifts describe the effect of charge density modifications on the action of individual gates, including the largest shifts in the closest gates and smaller, but still non-negligible shifts in farther gates due to crosstalk. The simulated pinch-off voltage shifts show good agreement with experimentally measured results, suggesting that gate-biased illumination results in a controllable, highly non-uniform charge distribution at the oxide-semiconductor interface. Finally, we measure the charge noise, an important metric for qubit operation, following a gate-biased illumination used to translate the quantum dots into the low-voltage operation regime. We find that there is no significant difference in charge noise when comparing the low-voltage operating regime to the typical operating regime attained after the device is illuminated with $0$~V applied to all its gates.

\section{Results and discussion}

\begin{figure}[ht!]
\centering
\includegraphics[width=3.3in]{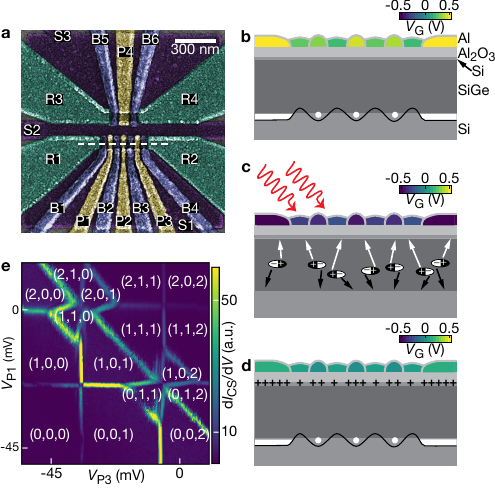}
\caption{\label{fig:fig1} Gate-biased illumination enables low-voltage operation of quantum dots. \textbf{a} A false-color SEM micrograph of a device that is nominally identical to the one measured here. \textbf{b}-\textbf{d} Cross-sections of the gate stack and upper layers of the Si/Si$_{0.7}$Ge$_{0.3}$ heterostructure along the white dashed line in (a) showing the gate-biased illumination process. \textbf{b} Large and non-uniform gate voltages are required to form a triple quantum dot due to fixed oxide charges and interface trapped charges (not shown). \textbf{c} The device is illuminated with a 780 nm laser while its gates are biased with dc voltages. Photo-generated electron-hole pairs (white-black ovals) move to screen the electric fields in the quantum well and are localized in traps. \textbf{d} After the gate-biased illumination, the newly trapped charges are frozen in place at the semiconductor-oxide interface, allowing the quantum dots to operate at smaller, more uniform gate voltages. \textbf{e} A stability diagram showing operation of a triple quantum dot in the (1,1,1) configuration at low gate voltages following the gate-biased illumination detailed in Table~\ref{tab:tab1}.}
\end{figure}

The measurements reported here are performed using a Si/SiGe triple quantum dot. Figure~\ref{fig:fig1}a is a false-colored scanning electron microscope (SEM) image of an overlapping-gate-defined semiconductor quantum dot device~\cite{Zajac2015-vu} with the same gate layout as the one measured here. Three layers of aluminum gates are patterned on top of a Si/Si$_{0.7}$Ge$_{0.3}$ heterostructure with a Al$_2$O$_3$ gate oxide layer~\cite{Dodson2020-jd}. The device is designed such that up to three quantum dots can be accumulated under plunger gates P1-P3, which are then measured by a charge sensor under the plunger gate P4. 

We adjust the operating voltages of the gates that define the quantum dots using a procedure we call gate-biased illumination, as illustrated in Fig.~\ref{fig:fig1}b-d. This is a generalization to multi-gate devices of the technique illustrated on Hall bars in Ref.~\cite{Wolfe2024-hs}. Figure~\ref{fig:fig1}b is a cross-sectional diagram of the gate stack and the top layers of the heterostructure, taken along the white dashed line in Fig.~\ref{fig:fig1}a. Here, the gates are shaded to depict the voltages that are applied when operating a triple quantum dot under the plunger gates P1-P3, which are enumerated in Table~\ref{tab:tab1}a. These voltages are large and non-uniform due to charge traps at the oxide-semiconductor interface and fixed charges in the oxide layer (not shown), of which many are intrinsic and some arise from imperfect fabrication. Additionally, gates usually must be set to different voltages depending on what purpose they serve. For instance, barrier gates are typically set to lower voltages than plunger gates to properly define quantum dots within the quantum well.

During the gate-biased illumination procedure, shown in Fig.~\ref{fig:fig1}c, the device is exposed to 780 nm light from a proximal diode laser while dc voltages are applied to its gates. Photons from the laser excite electron-hole pairs in the semiconductor heterostructure. As the heterostructure is saturated with electron-hole pairs, it temporarily behaves like a conductor, and the charge carriers rearrange to screen the Si quantum well from the electric fields generated by the gate voltages applied during illumination.

There is a large density of states at the oxide-semiconductor interface from the charge traps that exist there. The electrons or holes that screen the quantum well during the gate-biased illumination are confined in these charge traps, filling a portion of the density of states. Following the conclusion of the gate-biased illumination, these charge carriers are `frozen' in place, as shown in Fig.~\ref{fig:fig1}d. The voltages applied to the gates during the illumination determine the modification to the interface charge distribution, which directly influences the quantum dot operating voltages. By carefully selecting the voltages applied to the gates during illumination, it is possible to operate the device with smaller, more uniform gate voltages.

Figure~\ref{fig:fig1}e shows a stability diagram acquired after a gate-biased illumination with gate voltages set to the values shown in Table~\ref{tab:tab1}b. After illumination, we tune the triple quantum dot to the (1,1,1) charge configuration. We confirm that we are in this charge configuration by tuning the tunnel rates of the quantum dots such that each of the electron transitions around the (1,1,1) charge configuration is visible. The voltages applied to the gates in this low-voltage operating regime are listed in Table~\ref{tab:tab1}c, and in this configuration all the gate voltages are below 100 mV.

To determine how the uniformity of gate voltages changes after this gate-biased illumination, we compare the operating voltages listed in Table~\ref{tab:tab1}c to the typical operating voltages. These are defined as the gate voltages needed to tune the triple quantum dot to the (1,1,1) charge configuration following a gate-biased illumination with $0$~V applied to all the gates, and they are listed in Table~\ref{tab:tab1}a. We note an improvement in voltage uniformity in the low-voltage operating regime, with the difference between the largest gate voltage and the smallest gate voltage decreasing from 340 mV to 114 mV, a factor of 3 improvement.

\begin{table*}
\renewcommand{\thetable}{\arabic{table}}
\centering
\begin{tblr}{
  hlines,
  vlines = {dotted},
  vline{1-2,14} = {-}{solid},
  hline{3-4} = {-}{dotted},
}
Gate                                 & P1     & P2    & P3     & P4    & B1    & B2    & B3    & B4    & B5    & B6    & R1-R4 & S1-S3  \\
\textbf{a} $V_{\text{Typical}}$ (mV)        & 479  & 496 & 447  & 599.65 & 440  & 473 & 478 & 440  & 457 & 444 & 550  & 260  \\
\textbf{b} $V_{\text{Illumination}}$ (mV) & -500   & -500  & -500   & -500  & -300  & -300  & -300  & -300  & -400  & -400  & -500  & -200  \\
\textbf{c} $V_{\text{After}}$ (mV)      & -6 & 2.8 & -15 & 61 & 95 & 95 & 95 & 95 & 55 & 55 & 99 & 15 \\
\end{tblr}
\caption{\label{tab:tab1} Typical operating voltages (a) for the triple quantum dot in the single electron operation regime. Gate-biased illumination voltages (b) and operating voltages (c) for the low-voltage operation regime.}
\end{table*}

\begin{figure}[ht!]
\centering
\includegraphics[width=3.2in]{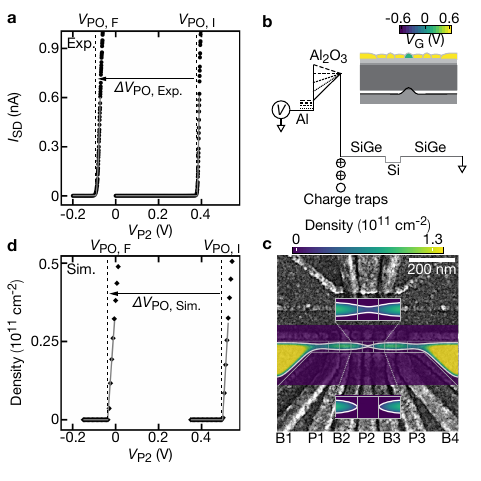}
\caption{\label{fig:fig2} Characterizing gate-biased illumination by simulating and measuring gate pinch-off voltage shifts. \textbf{a} Experimental pinch-off curves for the P2 plunger gate measured before and after illuminating the device with $\mathit{V}_{\mathrm{P2}}=-0.5~\mathrm{V}$. The pinch-off voltages from each measurement are indicated by the vertical dashed lines and the pinch-off voltage shift is the difference between the initial and final pinch-off voltages. The voltages applied to the gates during this measurement are shown in the inset of (b). \textbf{b} The model used to solve the charge density at the oxide-semiconductor interface during the gate-biased illumination. Here, accumulation occurs in charge-trap states at the interface depending on the voltage applied during illumination (dotted lines, voltages decrease from top to bottom). The change in interface charge density at different voltages is not shown. \textbf{c} The charge density in the quantum well during a simulation of the pinch-off voltage for P2. As the P2 gate is pinched off, the charge density in the channel goes from continuous (upper inset) to discontinuous (lower inset). \textbf{d} Simulated pinch-off curves for the P2 plunger gate before and after a simulated gate-biased illumination, using the charge density in the quantum well under the P2 gate as an analog to the current measured in (a). The pinch-off voltage shift is determined in the same way as in (a).}
\end{figure}

The key difference between the Hall bars studied in previous experiments~\cite{Saku1998-uy, Fujita2021-jn, Shetty2022-bl, Wolfe2024-hs, Stampfl2025-if} and quantum dot devices is that the latter have multiple nanoscale gates that can be set to different voltages. This means that to understand the mechanism behind gate-biased illumination in quantum dots, it is crucial to study gate-biased illumination in which the gates are held at non-uniform voltages. We do so by simulating the shift in pinch-off voltages after a gate-biased illumination and comparing it to experimental measurements of the same process. Figure~\ref{fig:fig2}a shows experimentally measured pinch-off curves for the P2 plunger gate following an illumination where $V_{P2}$ is held at $0$~V (right curve) and an illumination where $V_{P2}$ is set to $-0.5$~V (left curve). The voltages applied to the gates during the pinch-off measurements are shown in the inset of Fig.~\ref{fig:fig2}b. To extract the pinch-off voltage, the voltage at which the gate stops the flow of current between the two ohmic contacts, we fit the experimental data using the following piecewise function:

\begin{equation}
    I(V,V_{\mathrm{PO}},I_0,a,b) = \begin{cases}
        I_0 & V<V_{\mathrm{PO}} \\
        a(V-V_{\mathrm{PO}})^2+b(V-V_{\mathrm{PO}})+I_0 & V \geq V_{\mathrm{PO}}
    \end{cases}
    \label{eq:eq1}
\end{equation}

\noindent Here, $V_{PO}$ is the pinch-off voltage and $I_0$ is a small current offset from the measurement electronics. Since the information needed to extract the pinch-off voltage is fully contained in the low-current regime of these measurements, we fit to the data points that lie below a threshold of 0.6 nA, excluding all data points above the threshold from the fits.

Figure~\ref{fig:fig2}d shows simulated pinch-off curves of the P2 plunger gate following gate-biased illuminations with the same specifications as those in Fig.~\ref{fig:fig2}a. These simulations, as described below, solve for the charge density in the quantum well as a function of applied gate voltages, which we use as an analog for the current that flows in the quantum well during the experimental transport measurements. Using the same fitting function, but without thresholding, it is possible to extract gate pinch-off voltages from the charge density curves; however, due to a different global zero point in the simulations compared to the experimental measurements, it is not possible to directly compare the measured and simulated pinch-off voltages. Instead, we compare the shift in pinch-off voltages between the case in question and the base case of $0$~V applied to all the gates during illumination, since taking the difference in pinch-off voltages cancels this zero point.

As described above, the proposed physical mechanism responsible for the change in measured pinch-off voltages is the filling or emptying of charge traps at the oxide-semiconductor interface that occurs during a gate-biased illumination. To test this hypothesis, we simulate these effects. We first calculate the charge at the oxide-semiconductor interface needed to screen the Si quantum well from the electric fields generated by the gate voltages applied during illumination and then `freeze’ this charge in place, using it in a calculation of the resulting change in pinch-off voltages. If the resulting change in pinch-off voltages between the experiment and simulations match, this provides evidence that the physical picture described above is correct. To perform these simulations, we use the Modeling and Simulating for Quantum Exploration (MaSQE) codebase~\cite{Anderson2022-qi} in two steps. First, we use the 3D Laplace solver in MaSQE to treat the heterostructure as a bulk conductor during the illumination phase, and we determine the filling or emptying of charge traps at the oxide-semiconductor interface required to screen the electric field in the heterostructure, as shown in Fig.~\ref{fig:fig2}b. After computing the charge distribution needed to screen the electric field, we feed this distribution into a 3D self-consistent Schr{\"o}dinger-Poisson simulation, in MaSQE, where it acts as a `frozen’ charge distribution. We then calculate the electrostatic response of the device to changes in gate voltage and use this to estimate changes in pinch-off voltages. Figure~\ref{fig:fig2}c shows the charge density in the quantum well during a simulation used to determine the pinch-off voltage of the P2 gate. The insets show how the charge density changes as the voltage applied to the P2 gate is changed. As the P2 gate voltage is swept in the negative direction, the charge density goes from continuous (upper inset) to discontinuous (lower inset), indicating that the P2 plunger gate has pinched off the current channel. In the Supporting Information, we provide details of these simulations, including a description of the interface model used to provide the desired screening.

\begin{figure}[ht!]
\centering
\includegraphics[width=2.5in]{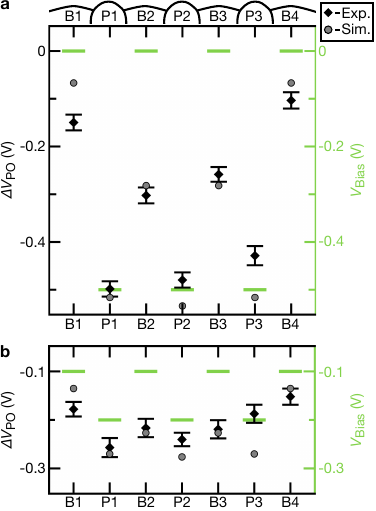}
\caption{\label{fig:fig3} Experimental (diamonds) and simulated (circles) shifts in pinch-off voltages showing good agreement for the following cases of gate-biased illumination: \textbf{a} case 1, large voltage difference between plunger and barrier gates ($-0.5$~V P1-P3 and $0$~V B1-B4), \textbf{b} case 2, smaller difference in plunger and barrier gate voltages ($-0.2$~V P1-P3 and $-0.1$~V B1-B4). For the experimental results, the data points are the mean pinch-off voltage shifts from 11 repetitions of each illumination and the error bars are the standard deviations. The green lines indicate the voltages applied to each gate during the illumination.}
\end{figure}

We now explore the effectiveness of gate-biased illumination by performing experiments and simulations in several different operating regimes. We select the following non-uniform voltage configurations that we apply to the device during illumination: (1) $-0.5$~V on plunger gates P1-P3, $0$~V on barrier gates B1-B4 and (2) $-0.2$~V on plunger gates P1-P3, $-0.1$~V on barrier gates B1-B4. In both cases, all other gates are set to $0$~V during illumination. In Fig.~\ref{fig:fig3}, we characterize the pinch-off voltage shifts following gate-biased illumination using the difference between each respective case and the base case of an illumination with $0$~V applied to each of the gates. For each case, including the base case, we repeat the illumination 11 times for robust statistics. The black diamonds in Fig.~\ref{fig:fig3} are the experimentally measured voltage shifts and the error bars are the standard deviations. The gray circles are the results of the simulations. Finally, the green lines indicate the voltages applied to each gate during the illumination.

Each case is designed to study an important aspect of implementing gate-biased illumination for the operation of a quantum dot device. The case where the plunger gates and the barrier gates have a large voltage difference (case 1) shows what happens when the plunger gates, which are often operated at larger voltages than their neighboring barrier gates, are biased by a very negative voltage during illumination. As shown in Fig.~\ref{fig:fig3}a, the pinch-off voltages for the plunger gates are shifted by a large amount in the negative direction. There are also sizable shifts in the pinch-off voltages of the barrier gates due to crosstalk from the neighboring plunger gates, as evidenced by the fact that the data points for the barrier gates do not line up well with the green lines indicating the voltages applied to these gates while illuminating the device. Notably, we see larger magnitude shifts in the barrier gates with two neighboring plunger gates (B2, B3) compared to barrier gates with only one neighboring plunger gate (B1, B4), as expected. The results of the simulations show good agreement with the experimental shifts, where they notably match the relative magnitude of the crosstalk effects on the barrier gates. 

In the case where the plunger and barrier gates have a smaller voltage difference (case 2), shown in Fig.~\ref{fig:fig3}b, we further study the effects of crosstalk by applying smaller, but non-uniform, voltages to the plunger gates and barrier gates during illumination. As before, we see shifts in the pinch-off voltages of the plunger and barrier gates in the negative direction. Here, all the pinch-off voltages are moderately shifted by crosstalk, such that the pinch-off voltage shifts for all the gates are of similar magnitude to the barrier gate shifts in Case 1, despite the fact that the applied voltages are significantly less negative in this case. This shows that we can achieve the voltage shifts necessary to translate the device into a low-voltage operating point without requiring excessively large negative voltages during illumination by taking advantage of the crosstalk effects. As in Case 1, there is reasonable agreement between the experimental shifts and the simulations.  The agreement between the experiment and the simulations in both Case 1 and Case 2 lends support for our model of charge being trapped primarily at the oxide-semiconductor interface during the gate-biased illumination process.

An advantage of gate-biased illumination is that it can be performed in under a minute, excluding the time spent to rethermalize the sample, making the characterization presented in Fig.~\ref{fig:fig3} much less time intensive compared to other methods that can be used to adjust quantum dot operating voltages. In this study, the vast majority of the time in each illumination cycle is spent waiting for the device to rethermalize after the illumination, in which is it heated up to $\sim 700$ mK. After each illumination, it takes $\sim 30$ minutes for the sample to cool back down below 100 mK, at which point we resume measuring the device. The sample heating is caused by Joule heating from the laser diode, which is positioned near the sample in the coldest part of the dilution refrigerator. As such, the waiting time for rethermalization can be effectively eliminated by positioning the laser farther from the sample, such as on the 4K plate, and delivering the near-infrared photons to the sample via an optical fiber.

\begin{figure}[ht!]
\centering
\includegraphics[width=2.95in]{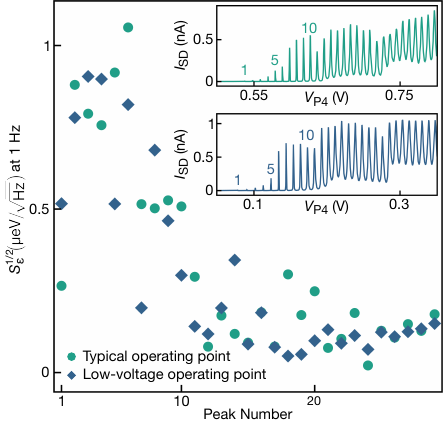}
\caption{\label{fig:fig4} Charge noise comparison between a typical operating regime and a low-voltage operating regime. \textit{Main panel}: charge noise $\text{S}_\text{\textepsilon}^{1/2}$ at 1 Hz for each of the first 29 charge sensor Coulomb peaks at a typical device operation point compared to the same peaks in a low-voltage operating regime following a gate-biased illumination. \textit{Inset panels}: the Coulomb peaks used when measuring the charge noise in each case. More details about the charge noise measurements can be found in the Supporting Information.}
\end{figure}

A potential concern with the gate-biased illumination procedure is that it could increase the charge noise seen by the quantum dots. This is because it results in a significant change in the electrostatic environment of the quantum dots, due to the amount of trapped charge introduced at the oxide-semiconductor interface. Thus, it is important to measure the charge noise to check whether qubit operations would be compromised by the process. To make a comparison between the charge noise present in a typical operating regime and the low-voltage operating regime attained using gate-biased illumination, we measure the charge noise of the first 29 Coulomb peaks (insets of Fig.~\ref{fig:fig4}) of the charge sensor under P4 in each case~\cite{Zwerver2022-vj, Paquelet-Wuetz2023-zc, Elsayed2024-os}. Here, the charge noise ($S_\varepsilon^{1/2}$) is defined as the fluctuations in the chemical potential of the charge sensor when it is tuned to the most sensitive point on each Coulomb peak. To acquire each charge noise spectrum, we average ten, ten-second time traces of the current through the charge sensor, taken 24 hours after the most recent illumination. These current measurements are converted into charge noise following the procedure outlined in the Supporting Information. Evaluating the charge noise at 1 Hz, as shown in the main panel of Fig.~\ref{fig:fig4}, it is reasonable to conclude that the charge noise does not increase after a gate-biased illumination, so we do not expect qubit operations to be negatively impacted. In both cases, the charge noise is comparable to values reported in the literature~\cite{Connors2019-wg, Paquelet-Wuetz2023-zc}.

\section{Conclusions}

Standardized tuning procedures that are necessary to scale to a large number of quantum dot qubits are made more complicated by non-uniformity in operating voltages for quantum dot gates. Here, we have shown that we can controllably, repeatably, and quickly tune the pinch-off voltages of a quantum dot device on a gate-by-gate basis by exposing it to near-infrared light while applying voltages to its gates. By comparing experimental measurements to results of self-consistent Schr{\"o}dinger-Poisson simulations, we lend support to a model of charge being trapped at the oxide-semiconductor interface during the gate-biased illumination process. Furthermore, we demonstrate an application of this technique by tuning the triple quantum dot to an operating regime with uniformly lower gate voltages while leaving the charge noise unchanged.

This work lays out a blueprint for the application of this technique to a large-scale quantum dot device. Following a basic characterization of the device and the pinch-off voltages for each of its gates, a set of voltages can be selected to translate it to a uniformly low-voltage operating regime. By combining information from basic measurements with simulations of the operating voltages, it may be possible to refine the voltages applied to the gates during illumination while requiring a minimal number of experimental measurements. Due to the short amount of time taken to perform a gate-biased illumination, this procedure could easily fit into the initial tuneup of a device~\cite{Kovach2026-ba}.  

Additionally, the generality of the mechanism behind gate-biased illumination means that it is relevant to a large class of gate-defined quantum dot devices. The key reason that this technique works is that there are stable trap states in the band gap of the semiconductor heterostructure. Any system that has such traps in the band gap is subject to non-uniformity and could be benefited by this method for tuning the operating voltages.

Larger-scale quantum dot devices will likely also rely on cryogenic voltage sources to supply their gate voltages. The power dissipated by these voltage sources is an important consideration, as it will need to be within the cooling budget of the cryostat that the device is mounted in. Reducing the operating voltages of the quantum dots using gate-biased illumination could help lessen this constraint, as smaller voltages could lead to lower power dissipation~\cite{Mouny2023-aq, Bian2024-hd}.

\section*{Data Availability Statement}

The datasets generated during the current study will be made available in a Zenodo repository prior to publication.

\section*{Acknowledgements}

The authors thank Chris Anderson for collaboration on the MaSQE simulations. The authors thank HRL for support and L. F. Edge of HRL Laboratories for providing the Si/SiGe heterostructure used in this work. Research was sponsored in part by the Army Research Office (ARO) under Grant Nos. W911NF-17-1-0274,  W911NF-22-1-0257, and W911NF-23-1-0110. The authors gratefully acknowledge the use of facilities and instrumentation in the Wisconsin Center for Nanoscale Technology. This Center is partially supported by the Wisconsin Materials Research Science and Engineering Center (NSF DMR-2309000) and by the University of Wisconsin–Madison. The views and conclusions contained in this document are those of the authors and should not be interpreted as representing the official policies, either expressed or implied, of the ARO, or the U.S. Government. The U.S. Government is authorized to reproduce and distribute reprints for Government purposes notwithstanding any copyright notation herein.

\section*{Competing Interests}
The authors declare no competing interests.

\section*{Supporting Information}

The Supporting Information includes the following:
\begin{itemize}
  \item Device and experimental setup information. Experimental methods including the gate-biased illumination process, photocurrent saturation measurements, and charge noise measurements. Details about the simulations.
\end{itemize}

\section*{Supporting Information for `Individually tunable Si/SiGe quantum dot operating voltages via gate-biased illumination'}

\setcounter{figure}{0}

\subsection*{Methods}

\subsubsection*{Device and heterostructure information}
The device measured in this study is a triple quantum dot fabricated with overlapping gates on a Si/Si$_{0.7}$Ge$_{0.3}$ heterostructure. The bottom (top) Si$_{0.7}$Ge$_{0.3}$ buffer layer is 170 nm (40 nm) thick and the Si quantum well is 9 nm thick. The heterostructure is capped with a 1 nm layer of Si. The Al$_2$O$_3$ gate oxide is 5 nm thick and fabricated using atomic layer deposition. The native AlO$_x$ intergate oxide is grown using plasma ashing and is 4 nm thick~\cite{Dodson2020-jd}. The gate layer thicknesses for the screening gates, accumulation gates, and barrier gates are 30 nm, 50 nm, and 65 nm, respectively.

\subsubsection*{Experimental setup}
The experiment is performed in a wet dilution refrigerator with a base temperature of $\lesssim 30$ mK. The device gates are biased using SRS SIM928 isolated voltage sources. Source-drain currents measured between the ohmic contacts are amplified at room temperature using a DL Instruments Model 1211 transimpedence preamplifier and isolated from digital electronics using an SRS SR560 voltage preamplifier. DC measurements are digitized using a National Instruments NI-USB 6216 and lock-in measurements are performed by two Signal Recovery 7265 lock-in amplifiers. The diode laser used for illumination is powered using a Keithley 2400 Source Measure Unit.

\subsubsection*{Gate-biased illumination}
To perform a gate-biased illumination, the device is illuminated for 40 seconds with a 780 nm diode laser while voltages are applied to its gates. The specific laser used is a US-Lasers D780-5, which is a 5 mW laser that we operate at a constant current of 20 mA supplied by a Keithley 2400 Source Measure Unit. During an illumination, the temperature of the mixing chamber of the dilution refrigerator typically increases to ~0.7K - 1K due to Joule heating of the laser diode and takes about half an hour to rethermalize. We ensure that the gate voltage shifts induced by gate-biased illumination are caused by this process and not any bias cooling effects by setting any nonzero gate voltages back to zero once we stop shining the laser on the device. The time spent illuminating the device could be decreased through better alignment with the laser diode, and the time to rethermalize could be eliminated by moving the laser farther from the mixing chamber, through the use of a fiber-coupled laser mounted to a higher temperature plate in the dilution refrigerator.

\subsection*{Photocurrent saturation}

\begin{figure}[ht!]
\centering
\renewcommand{\thefigure}{S\arabic{figure}}
\includegraphics[width=3.3in]{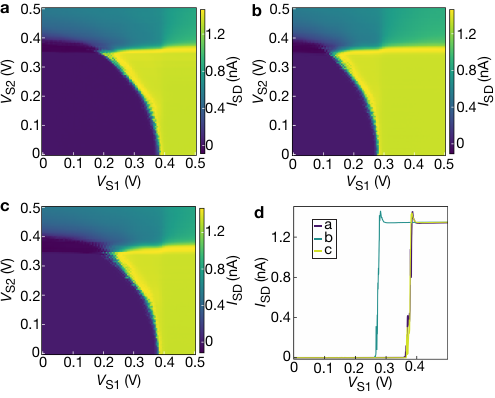}
\caption{\label{fig:figS1} Saturation of photocurrent during the illumination process. \textbf{a} Triangle plot following an illumination with zero applied gate voltages. \textbf{b} Triangle plot following an illumination with $-0.1$~V applied to the screening gate S1 and $0$~V applied to the other gates. \textbf{c} Triangle plot following a second illumination with zero applied gate voltages. \textbf{d} Linecuts from each of the triangle plots at $V_{S2}=0~\mathrm{V}$ showing that the triangle plot shifts by an appropriate amount in each step of the photocurrent saturation check.}
\end{figure}

For the results of the gate-biased illumination experiment to be consistent, it is important that the semiconductor heterostructure be sufficiently saturated with electron-hole pairs so that the applied biases on the device gates can be screened identically in each repetition of the experiment. Otherwise, the charge distribution at the interface may vary from illumination to illumination, resulting in inconsistent voltage shifts. To show that the parameters of the gate-biased illuminations used in this study, as described above, are sufficient to produce repeatable results, we perform a simple photocurrent saturation check. First, we illuminate the device with $0$~V applied to all the gates and measure a triangle plot, as shown in Fig.~\ref{fig:figS1}a. This measurement characterizes the action of the screening gates on the one-dimensional (1D) current channel that is formed between them. Next, we illuminate with $-0.1$~V applied to the screening gate S1 and measure the same triangle plot again. Figure~\ref{fig:figS1}b shows this measurement, which has a clear shift in the negative direction in $V_\mathrm{S1}$. Finally, we repeat the first illumination with $0$~V applied to all the gates and check that the triangle plot shifts back to the original position in voltage space in Fig.~\ref{fig:figS1}c. Figure~\ref{fig:figS1}d shows linecuts from all three triangle plots, clearly illustrating that the high-current region in the triangle plot shifts by $\sim-100~\mathrm{mV}$ after the gate-biased illumination on S1 before shifting back to its original position after the final illumination.

\subsection*{Charge noise measurements}
\begin{figure}[ht!]
\centering
\renewcommand{\thefigure}{S\arabic{figure}}
\includegraphics[width=3.3in]{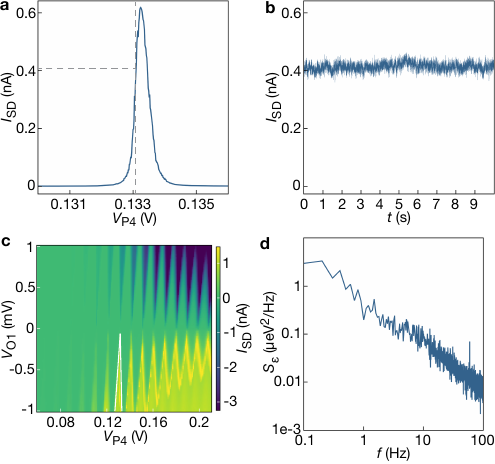}
\caption{\label{fig:figS2} Charge noise measurements following a gate-biased illumination. \textbf{a} A typical charge sensor peak used for charge noise measurements. \textbf{b} A time trace measured by sitting on the flank of the charge sensor peak. \textbf{c} Coulomb diamond measurements are used to calculate the lever arm of the charge sensor. \textbf{d} The charge noise spectrum is calculated by averaging the fast Fourier transforms of ten, ten second time traces and converting from voltage noise to chemical potential noise using the sensitivity of the charge sensor and the lever arm.}
\end{figure}

Figure~\ref{fig:figS2} shows the methodology behind the charge noise measurements presented in Fig.~\ref{fig:fig4} of the main text. Here, we measure charge noise by parking the charge sensor at the gate voltage that maximizes its sensitivity, as shown in Fig.~\ref{fig:figS2}a. At this sensitive point, we measure the current fluctuations over time. Figure~\ref{fig:figS2}b shows an example of a time trace of the charge sensor current. We measure ten second time traces, which we then convert to the frequency domain using a fast Fourier transform. To convert from current noise to voltage noise, we use the sensitivity of the charge sensor peak. We measure the lever arm of the sensor dot by measuring Coulomb diamonds in transport. We extract the lever arms from these transport measurements using the same method as in Ref.~\cite{Paquelet-Wuetz2023-zc}, and the white lines in Fig.~\ref{fig:figS2}c are an example of the lines whose slopes we use to calculate the lever arm for the charge sensor for one of its Coulomb peaks. Using this lever arm, we convert the spectrum from voltage noise to chemical potential noise ($S_\varepsilon$). Finally, we arrive at the spectrum shown in Fig.~\ref{fig:figS2}d by averaging ten spectra that are calculated from consecutively measured time traces.

\subsection*{Simulating the gate-biased illumination process}

As described in the main text, the aim of the simulations is to calculate the change in pinch-off voltages due to the rearrangement of charge at the interface that occurs during gate-biased illumination. In this process, the heterostructure temporarily behaves like a conductor and charges rearrange to screen the Si quantum well from the electric fields generated by the gate voltages. The three-dimensional (3D) physical model used here is motivated by the 1D biased illumination Hall bar work described in Ref.~\cite{Wolfe2024-hs}. We use the Modeling and Simulating for Quantum Exploration (MaSQE) codebase~\cite{Anderson2022-qi} to determine the filling or depletion of charge traps during gate-biased illumination by modeling the heterostructure as a conductor for the time period that the heterostructure is under illumination (see Fig.~\ref{fig:fig2}b of the main text). Using this model, the simulation has two steps. The first step uses MaSQE to simulate screening effects caused by gate-biased illumination, in 3D, in response to realistic gate voltages used in the experiments. The second step `freezes' in the resulting non-uniform two-dimensional (2D) sheet charge density to calculate the change in pinch-off voltages for each finger gate.

To simulate gate-biased illumination in three dimensions, we perform 3D Laplace simulations in MaSQE. We first define a heterostructure that contains a single 5 nm thick Al$_2$O$_3$ dielectric layer, and then we pattern the surface of the device with overlapping gates, matching the SEM image shown in Fig.~\ref{fig:fig1}a of the main text. The experimental voltages applied to the overlapping gates, as described in the main text, are imposed as Dirichlet boundary conditions at the top of the device. We impose a lower Dirichlet boundary condition of $V = 0$~V so that holes fill the charge traps when applying a negative top gate bias, and electrons fill the charge traps when applying a positive top gate bias. We assume here that the experiments are performed within the linear regime for charge-trap filling, as discussed in Ref.~\cite{Wolfe2024-hs}. We use MaSQE to solve for the perpendicular electric field $E_{z}(x,y)$ incident on the lower interface and use Eq.~\ref{eq:charge} to determine the spatially varying charge $\sigma(x,y)$ at the oxide-semiconductor interface needed to screen $E_{z}(x,y)$:

\begin{equation}\label{eq:charge}
    \sigma(x,y) = \varepsilon_{\mathrm{Al_2O_3}} \varepsilon_{0} E_{z}(x,y).
\end{equation}

To calculate the change in pinch-off voltages for each finger gate, we introduce the non-uniform 2D charge distribution computed above, which is now frozen in place at the oxide-semiconductor interface, and perform 3D self-consistent Schr{\"o}dinger-Poisson simulations in MaSQE. To mimic experimental conditions, we apply a uniform voltage shift to all gates that is appropriate to accumulate a charge of about $2\times 10^{11}$ e/cm$^2$ in the reservoirs. To this uniform voltage shift, we add the voltages applied to every gate in the pinch-off experiments, resulting in a 1D channel of accumulated charge in the silicon quantum well of the device, similar to the top inset of Fig.~\ref{fig:fig2}c of the main text. For each gate, the pinch-off voltage is determined by sweeping that gate's voltage and identifying the voltage that depletes the charge density in the quantum well directly beneath the gate, as described in the main text. Figure~\ref{fig:fig2}c shows how the charge density in the channel changes from continuous (upper inset) to pinch-off (main panel) to discontinuous (lower inset) as the voltage applied to the P2 gate is swept in the negative direction.

As a final technical note, the MaSQE Schr{\"o}dinger-Poisson solver, as used here, treats the quantum well quantum mechanically in the $z$ direction and classically in the $x$ and $y$ directions. To account for the quantum confinement in the lateral $y$ direction, perpendicular to the 1D channel during pinch-off, we calculate the first $y$ subband energy at the position of the minimum charge density (where the pinch-off occurs) and then adjust the voltage on the given finger gate to shift the potential so that the first $y$ subband lies just above the Fermi level. We note that this small correction is largely canceled out when comparing the difference in pinch-off voltages between different bias configurations.

\newpage

\printbibliography

\end{document}